\documentclass[aps,prd,amssymb,showpacs,floatfix,twocolumn,superscriptaddress,nofootinbib]{revtex4-1}
\usepackage{amsmath,amsfonts,graphics,epsfig}

\begin{document}

\title{Quadrupole moments of
spin--1 systems:
the $\rho$ meson, the $S$--wave deuteron
and some general constraints}
\author{A.F.~Krutov}
\email{krutov@ssau.ru}, \affiliation{Samara University, 443086 Samara,
Russia}, \affiliation{Samara State Technical University, 443100 Samara,
Russia}\author{V.E.~Troitsky} \email{troitsky@theory.sinp.msu.ru}
\affiliation{D.V.~Skobeltsyn
Institute of Nuclear Physics,\\
M. V. Lomonosov Moscow State University, Moscow 119991, Russia}
\date{\today}
\begin{abstract}

We costruct the relativistic operator of the quadrupole moment of
two-particle composite spin one systems with zero orbital moment of
the relative motion and derive explicit analytical expression for the
quadrupole moment
using the approach to relativistic composite systems based on our version
of the instant-form relativistic quantum mechanics (RQM). We calculate the
quadrupole moments of the $\rho $ meson and of the $S$-wave deuteron
without any free parameters, using our unified $\pi $\&$\rho $ model
(Phys. Rev.D \textbf{93}, 036007 (2016); \textbf{97}, 033007 (2018))
and our previous results on deuteron. Our calculation gives
$Q_{\rho} = -0.158\pm 0.04$ GeV$^{-2}$ and
$Q_{d} = -1.4\cdot 10^{-4}$ GeV$^{-2}$.
Having in our
disposition the rather general form of the quadrupole-moment operator we
for the first time formulate the problem of the upper and lower bounds for
possible values of the quadrupole moment of a two-particle system with
indicated quantum numbers for a large range of constituent masses, and
partially solve it.

\end{abstract}

\maketitle

\section{Introduction}
\label{sec: introduc}

The electroweak properties of hadrons (decay constants, mean square radii,
static moments, electromagnetic form factors etc.) are of  fundamental
importance for the understanding of strong interactions at low and
intermediate energy scales. So, it is clear that the theory of such
properties based on different nonperturbative approaches is in the focus
of investigations for years. Let us mention for example different forms of
Dirac relativistic dynamics
\cite{BaE85, CaG95, ChC88, MeF97, CaD98, BaC02, Jau03, ChJ04, HeJ04,
BiS14, MeS15, SuD17},
approaches based on the
Dyson-Schwinger equation \cite{HaP99, BhM08, RoB11, PiS13}, the
Nambu-Jona-Lasinio model \cite{CaB15, LuC15},
QCD sum rules\cite{Sam03, AlO09}, the light front diagram technique
\cite{MeS02}, lattice calculations
\cite{AnW97, HeK07, OwK15, LuS16}.
The quadrupole moment of two-particle spin one systems with total electric
charge equal to unity with $S$-state of the relative motion of constituents
is of particular interest because the existence of the quadrupole form
factor and the quadrupole moment in such systems is a purely relativistic
effect. They are strictly zero in the non-relativistic case. The actual
cause of the effect is well known (see, e.g., the textbook \cite{Nov75}):
the relativistic spin rotation of the constituents. This effect is a
kinematic one, and so it must show itself in all composite systems with
given quantum numbers, being independent from the nature of constituents
and the kind of their interaction. In particular, this effect takes place
in two well studied systems with principally different constituents and
types of the interaction: the $\rho$ meson and the deuteron. It is worth
noting that a kind of universal conditions for the quadrupole moment of
such systems was given in the well known paper \cite{BrH92}.

The results of calculations of $\rho $-meson the quadrupole moment
through different approaches (see, e.g., ~\cite{BaE85, CaG95, MeF97,
CaD98, BaC02, Jau03, ChJ04, HeJ04, BiS14, CaB15, MeS02})
differ essentially not only in the absolute value but even in sign.
To-day it is not possible to estimate the credibility of these
results because
the experimental data on the $\rho $ meson are scarce.
Its
lifetime is very short, so direct
measurements of its electroweak properties are nearly impossible.

Although the deuteron quadrupole moment is measured with great
accuracy (see, e.g.,
\cite{GaO01, GiG02}), it is clear that the quadrupole form factor and the
quadrupole moment of the deuteron are mainly defined by the
$D$-wave part of the deuteron wave function.
The $S$-state admixture is small and it is very difficult
to separate it. The contribution of the pure $S$-state to the
 quadrupole moment is relatively small and it is doubtful whether it
 can be extracted from the experiment.

The goal of the present paper is threefold. First, we derive
an explicit analytic formula for the quadrupole moment
$Q$ of a two-particle spin one system in
the $S$-state of relative motion of constituents. Second, we calculate
the quadrupole moments of
the $\rho$ meson and the $S$-wave deuteron using no free parameters.
Third, in the framework of our relativistic approach we obtain some
general constraints for possible values of the quadrupole moment in
two-particle systems with quantum numbers given above for a large range
of constituent masses from $\rho $ meson to deuteron.

The approach that we use in the present paper is a particular
relativistic formulation
of constituent-quark model that is based on the classical paper
by P.~Dirac \cite{Dir49} (so-called Relativistic Hamiltonian
Dynamics or
Relativistic  Quantum Mechanics (RQM)). RQM can be formulated in different
ways or in different forms of dynamics. The main forms are instant form
(IF), point form (PF) and light-front (LF) dynamics. The properties of
different forms of RQM dynamics can be found in
the reviews \cite{LeS78, KeP91, Coe92, KrT09}. Today the approach is
largely used for nonperturbative theory of particle structure.

Here we use our version of
RQM, -- the modified instant form (mIF) of RQM ,
that was successfully used for various composite two-particle systems,
namely, the deuteron
\cite{KrT07}, the pion
\cite{KrT01, KrT98, KrT09prc, TrT13}, the $\rho$ meson \cite{KrP16, KrP18}
and the kaon \cite{KrT17}. This model has predicted with surprising
accuracy the values of the form factor $F_{\pi}(Q^{2})$, which were
measured later in JLab experiments \cite{Vol01, Hor06, Tad07, Blo08,
Hub08} (see discussion in Ref. \cite{KrT09prc}). All the new points fall,
within the experimental uncertainties, on the initially calculated curve.
Another advantage of the approach is matching with the QCD predictions in
the ultraviolet limit: when constituent-quark masses are switched off, as
expected at high energies, the model reproduces correctly not only the
functional form of the QCD asymptotics, but also its numerical
coefficient; see Refs. \cite{KrT98, TrT13, TrT15} for details. The method
allows for an analytic continuation of the pion electromagnetic form factor
from the spacelike region to the complex plane of momentum transfers and
gives good results for the pion form factor in the timelike region
\cite{KrN13}.

Now we use this approach, supplemented with some physical reasoning based
on the consideration of the structure of our relativistic operator of the
quadrupole moment, to obtain some constraints for possible values of the
quadrupole moment of two-particle spin one systems in the $S$-state of
relative motion. The effectiveness of our approach for a relativistic
theory of two-particle composite systems permits us to believe that our
constraints are of a rather general character.

In what concerns the calculation of the quadrupole moment of the $\rho $
meson, the present paper is a continuation of
our papers \cite{KrP16, KrP18}. Using our  approach we have constructed
\cite{KrP16} the unified $\pi $\&$\rho $ model and have fixed all free
parameters determining the radius of the $\rho$ meson through its decay
constant. Then we obtain the $\rho$ meson magnetic moment $\mu_{\rho}$
\cite{KrP18} using the unified $\pi $\&$\rho $ model with no new fitting
parameters. Now we calculate the $\rho $-meson quadrupole moment in the
unified $\pi $\&$\rho $ model, that is without any fitting parameters: all
parameters are already fixed in \cite{KrP16}.

The reliability of our calculation of the quadrupole moment of the
$S$-wave deuteron in our approach is based on the results of the paper
\cite{KrT07} where very good theoretical results were derived
for the electromagnetic properties of the deuteron obtained in polarization
experiment on the electron-deuteron scattering, that is for the component
$T_{20}(Q^2)$ of the deuteron polarization tensor and for its
quadrupole form factor. In the present paper we use
the $S$-component of the deuteron wave function
\cite{KrT07MT}, that was constructed in the framework of the so-called
potentialless formulation of the inverse scattering problem
\cite{MuT81} (see also \cite{Tro94}).

Our relativistic operator of the quadrupole moment of the two-particle
system constructed in the basis of state-vectors with the motion of
center-of-mass being separated is a c-number function. To obtain some
general results for arbitrary constituents with spin 1/2 in the $S$-state
of relative motion we divert our attention away from the fixed parameters
in unified $\pi $\&$\rho $ model or in the deuteron wave function.
Now we
consider the operator of the quadrupole moment on the  class of
parameters that characterize, on the one hand, the constituents (the mass,
the anomalous magnetic moments) and, on the other hand, the interaction of
constituents. We consider weak interactions (as in deuteron),
intermediate model interaction (see, e.g., \cite{KrT01, KrP18})
and strong interactions  that ensure square-law quark confinement
(harmonic oscillator wave function).

The derived expression for the relativistic quadrupole-moment operator
suggests that one can obtain some general limitations for the values of
quadrupole moment of systems under consideration. The analysis of the
formula for the quadrupole-moment operator, some additional physical
reasoning and involvement of numerical calculations enable us to construct
the upper and lower bounds for the values of quadrupole moments under
consideration, resulting in some constraints.

The rest of the paper is organized as follows. In Sec.
\ref{sec: Quad s-st}, the  quadrupole form factor and the quadrupole moment
of a spin one two-particle system in the  $S$-state of the relative motion
are derived in modified impulse approximation of IF RQM. In Sec. \ref{sec:
Quad rho deut} we calculate the  values of quadrupole moments of the $\rho
$ meson (using the unified $\pi $\&$\rho $ model with no free parameters)
and of the deuteron in $S$-state with no $D$-wave admixture. Sec.
\ref{sec: Properties Q} contains the analysis of the  properties of the
relativistic quadrupole-moment operator for indicated quantum numbers and
of the dependence of its value on the model interaction of constituents as
well as on the values of constituent masses and anomalous magnetic
moments. In Sec. \ref{sec: Quad constr} general limitations for  the
values of the quadrupole moment of different systems with mentioned
quantum numbers are proposed and discussed. We briefly conclude in
Sec.\ref{sec:concl} and present some details of the calculation in the
Appendix.

\section{The quadrupole moment of $S$-state two-particle\\
spin one system as a relativistic effect}
\label{sec: Quad s-st}

One of  the main points of our approach is the construction of matrix
element of electromagnetic current for a composite system of two
interacting particles. A summary of our method of such construction can be
found in our recent paper \cite{KrP18} and in the references therein. The
method is based on the principal statements of RQM dynamics(see, e.g.,
\cite{BaT53})
and on  the general procedure of relativistic covariant construction
of local operators matrix elements \cite{ChS63}.

Let us consider a system of two interacting particles of the mass $M$, the
spin 1/2 and the total electric charge 1 in the $S$-state of relative
motion. In RQM the basis of individual spins and momenta of particles can
be used
\begin{equation}
\left|\,\vec p_1\,,m_1;\vec p_2\,,m_2\rangle\right. =  \left|\,\vec
p_1\,,m_1\rangle\otimes|\vec p_2\,,m_2\rangle\right.\;,
\label{p1m1p2m2}
\end{equation}
where $\vec p_{1,2}$ are the constituent momenta and $m_{1,2}$ are their
spin projections.
One can also choose the following set of two-particle state vectors
where the motion of the
center of mass is separated:
\begin{equation}
|\,\vec P,\;\sqrt {s},\;J,\;l,\;S,\;m_J\,\rangle\;,
\label{PkJlSm}
\end{equation}
where $P_\mu = (p_1 +p_2)_\mu$, $P^2_\mu = s$; $\sqrt {s}$
is the invariant
mass of the two-particle system, $l$ is the orbital angular
momentum in the center-of-mass frame (C.M.S.),
$\vec S\,^2=(\vec S_1 +\vec S_2)^2 = S(S+1)\;,\;S$
is the total spin in
C.M.S., $J$ is the total angular momentum with the projection
$m_J$.
The two-particle basis with separated motion of the center of mass
(\ref{PkJlSm}) is connected with the basis of individual spins and
momenta of two particles (\ref{p1m1p2m2})
through the appropriate
Clebsh-Gordan  decomposition
 for the Poincar\'e group
(see, e.g., \cite{KrT09}).

The  current matrix element for our system is
\begin{equation}
\langle\vec p_c\,,m_{Jc}|j_\mu(0)|\vec p_c\,'\,,m'_{Jc}\rangle\;,
\label{jc}
\end{equation}
where $\vec p_c\;,\;\vec p_c\,'$ are the momenta of composite
two-particle system
in initial
and final states, respectively,
$m_{Jc}\;,m'_{Jc}$ are
projections of the total angular momenta.
As the  expression (\ref{jc}) is a matrix in the projections of the total
angular momentum, it can be decomposed in the sum of linearly independent
matrices (see for detail \cite{KrT09, KrP18, ChS63}) that presents a set
of
$2J+1$ independent Lorentz scalars (that is scalars or
pseudoscalars):
\begin{equation}
D^{J_c}(p_c,p'_c)\left(p_{c\mu}\Gamma^\mu\left(p'_c\right)\right)^n\;,\quad n = 0,1,2,\ldots,2J\;,
\label{scalars}
\end{equation}
here $D^J$
is the matrix of Wigner rotation (see, e.g.,
 \cite{Nov75}).
The spin 4-vector
$\Gamma^\mu$ \cite{KrT09, ChS63} is defined as follows:
$$
\Gamma_0(p_c) = (\vec p_c\vec J)\;,\quad \vec \Gamma(p_c)
= M_c\,\vec J_c + \frac {\vec p_c(\vec p_c\vec
J_c)}{{p_c}_0 + M_c}\;,
$$
\begin{equation}
\quad \Gamma^2 = -M_c^2\,J_c(J_c+1)\;. \label{ Gamma mu}
\end{equation}

In the decomposition of (\ref{jc}) in terms of the set (\ref{scalars})
each Lorentz scalar is multiplied by a 4-vector constructed of variables
that enter the state vectors in initial and finite states. So, the
decomposition has the form (see also \cite{KrP18, KrT03}):
$$
\langle\vec p_c\,,\,m_{Jc}|j_\mu(0)|\vec
p_c\,'\,,\,m'_{Jc}\rangle =
$$
$$
\langle\,m_{Jc}|\,D^{1}(p_c\,,\,p'_c)\, \sum_{i=1,3}\,
\tilde{\cal F}\,^i_c(t)\,\tilde A^i_\mu\,|m'_{Jc}\rangle\;,
$$
$$
\tilde{\cal F}\,^1_c(t) = \tilde f^c_{10} + \tilde
f^c_{12}\left\{[i{p_c}_\nu\,\Gamma^\nu(p'_c)]^2 \right.
\left. \right. -
$$
\begin{equation}
\left.
\frac{1}{3}\,\hbox{Sp}[i{p_c}_\nu\,\Gamma^\nu(p'_c)]^2\right\}
\frac{2}{\hbox{Sp}[{p_c}_\nu\,\Gamma^\nu(p'_c)]^2}\;,
\label{fin}
\end{equation}
$$
\tilde{\cal F}\,^3_c(t) = \tilde f^c_{30}\;,
$$
$$
\tilde A^1_\mu = (p_c + p'_c)_\mu\;,\quad \tilde A^3_\mu =
\frac{i}{M_c} \varepsilon_{\mu\nu\lambda\sigma}
\,p_c^\nu\,p'_c\,^\lambda\,\Gamma^\sigma(p'_c)\;.
$$
Here $M_c$
is the mass of composite system, $Q^2 = -q^2 = t$, $q$ - 4-vector of
the momentum transfer, $\tilde f^c_{10}\,,\,\tilde f^c_{12}\,,$ $\tilde
f^c_{30}$ are the charge, quadrupole and magnetic form factors,
respectively.

The invariant parts that one can extract from the matrix element
are called the Sachs form factors of the composite system. One have the
charge $G_C(Q^2)$, quadrupole $G_Q(Q^2)$ and magnetic $G_M(Q^2)$ form
factors (see, e.g., \cite{BrH92, ArC80}). The Sachs form factors can be
written in terms of the form factors in (\ref{fin}) as follows:
$$
G_C(Q^2) = \tilde f^c_{10}(Q^2)\;,\quad G_Q(Q^2) =
\frac{2\,M_c^2}{Q^2}\,\tilde f^c_{12}(Q^2)\;,
$$
\begin{equation}
\quad G_M(Q^2) = -\,M_c\,\tilde f^c_{30}(Q^2)\;.
\label{G=f}
\end{equation}

The current matrix element in RQM (\ref{jc}) can be decomposed in the
 complete set of states (\ref{PkJlSm}):
$$
\langle\vec p_c\,,\,m_{Jc}|j_\mu(0)|\vec p_c\,'\,,\,m'_{Jc}\rangle =
$$
$$
\sum\limits_{m_J}\int\frac{d\vec P\,d\vec P\,'}{N_{CG}N'_{CG}}d\sqrt{s}\,d\sqrt{s'}
\langle\vec p_c\,,\,m_{Jc}|\,\vec P\,,\,\sqrt {s}\,,\,m_J\,\rangle
$$
$$
\langle\vec P\,,\,\sqrt {s}\,,\,m_J\,|j_\mu(0)|\vec P\,'\,,\,\sqrt {s'}\,,\,m'_J\,\rangle
$$
\begin{equation}
\langle\,\vec P\,'\,,\,\sqrt {s'}\,,\,m'_J|\vec p_c\,'\,,\,m'_{Jc}\rangle\;,
\label{jcDecom}
\end{equation}
We do not use in the present paper the explicit form of the  normalization
constant $N_{CG}$ of the vectors (\ref{PkJlSm})  (it can be found in
\cite{KrT09}); $\langle\,\vec P,\sqrt {s}, m_J|\vec
p_c,m_{Jc}\rangle$ is the wave function  of the
composite system in the sense of RQM in the representation defined by the
basis
 (\ref{PkJlSm}). In the state vectors
in (\ref{PkJlSm}) the fixed quantum numbers $J = S = 1\;,\;l=0$ are
omitted.

The wave function of the composite system in (\ref{jcDecom}) is:
\begin{equation}
\langle\,\vec P\,,\,\sqrt {s}\,,\,m_J|\vec p_c\,,\,m_{Jc}\rangle = N_C\,\delta(\vec P - \vec p_c)\,\delta_{m_J m_{Jc}}\varphi(s)\;,
\label{wf}
\end{equation}
where $N_C$ is the normalization constant (see \cite{KrT09}) that we do
not need here.

The wave function of the relative motion
$\varphi(s)$  in the representation defined by the basis (\ref{PkJlSm}) for
$l=0\,,\,S=1$
 is a solution of the
eigenvalue problem for the mass (or the mass square) operator for
two-particle system with interaction:
$\hat{M_I} = \hat{M_0} + \hat{V}$, where $\hat{M_0}$ is the mass operator
for two-particle system without interaction and $\hat{V}$ is the
interaction operator.
The wave function has the form
\begin{equation}
\varphi(s) = \sqrt[4]{s}\,k\,u(k)\;,\quad s = 4(k^2 + M^2)\;,\quad \int\limits_0^\infty dk\,k^2\,u^2(k) = 1\;,
\label{phi}
\end{equation}
with $M$ being the individual mass of a constituent.

Taking into account (\ref{wf}) we rewrite the decomposition (\ref{jcDecom})
in the form:
$$
\langle\vec p_c\,,\,m_{Jc}|j_\mu(0)|\vec p_c\,'\,,\,m'_{Jc}\rangle =
$$
$$
\int\frac{N_C\,N'_C}{N_{CG}\,N'_{CG}}d\sqrt{s}\,d\sqrt{s'}\varphi(s)
$$
\begin{equation}
\langle\vec p_c\,,\,\sqrt {s}\,,\,m_{Jc}|j_\mu(0)|\vec p_c\,'\,,\,\sqrt {s'}\,,\,m'_{Jc}\,\rangle\,\varphi(s')
\;.
\label{jcRed}
\end{equation}
The matrix element in (\ref{jcRed})
\begin{equation}
\frac{N_C\,N'_C}{N_{CG}N'_{CG}}\langle\vec p_c\,,\,\sqrt {s}\,,\,m_{Jc}|j_\mu(0)|\vec p_c\,'\,,\,\sqrt {s'}\,,\,m'_{Jc}\,\rangle
\label{GenF}
\end{equation}
is a regular Lorentz-covariant generalized function (distribution) that
has a meaning only under the integral in
 (\ref{jcRed}). So, the integral (\ref{jcRed}) is to be
 regarded as a functional
defined on the space of test functions
$\varphi(s)\varphi(s')$.

Now we decompose the matrix element  (\ref{GenF}) in the r.h.s.of
(\ref{jcRed}) in the system of independent Lorentz-scalars (\ref{scalars})
in analogy to (\ref{fin}):
$$
\frac{N_C\,N'_C}{N_{CG}N'_{CG}}\langle\vec p_c\,,\,\sqrt {s}\,,\,m_{Jc}|j_\mu(0)|\vec p_c\,'\,,\,\sqrt {s'}\,,\,m'_{Jc}\,\rangle =
$$
$$
\langle\,m_{Jc}|\,D^{1}(p_c\,,\,p'_c)\,\sum_{i=1,3}\,
{\cal F}\,^i(s,Q^2,s')\,B^i_\mu(s,Q^2,s')\,|m'_{Jc}\,\rangle\;,
$$
$$
{\cal F}\,^1(s,Q^2,s') = G_{10}(s,Q^2,s') +
$$
$$
G_{12}(s,Q^2,s')\left\{[i{p_c}_\nu\,\Gamma^\nu(p'_c)]^2 \right.
\left. \right. -
$$
\begin{equation}
\left.
\frac{1}{3}\,\hbox{Sp}[i{p_c}_\nu\,\Gamma^\nu(p'_c)]^2\right\}
\frac{2}{\hbox{Sp}[{p_c}_\nu\,\Gamma^\nu(p'_c)]^2}\;,
\label{GenG}
\end{equation}
$$
{\cal F}\,^3(s,Q^2,s') = G_{30}(s,t,s')\;,
$$
where $B^i_\mu(s,t,s')\;,\;i=1,3$ are some 4-vectors that are smooth
functions of the variables $s,s'$ .

Substituting of the decompositions (\ref{fin}),
(\ref{GenG}) in (\ref{jcRed}) and equating the expressions that stand at
the equal degrees of the scalars
(\ref{scalars}) we obtain some equalities for the 4-vectors.
These equalities are to be hold in the sense of
Lorentz-covariant
generalized functions, that is for arbitrary test functions
$\varphi(s)\varphi(s')$.
This condition
means that these covariant relations are to be valid for arbitrary model
of the interaction of constituents in RQM. So. the vectors
$B^i_\mu(s,t,s')\;,\;i=1,3$  in
(\ref{GenG}) are the same as $\tilde{A^i_\mu}$ in (\ref{fin}).
As a result we obtain for the invariant parts of the matrix element
(\ref{jc}):
\begin{equation}
\tilde f^c_{in}(Q^2) =  \int\,d\sqrt{s}\,d\sqrt{s'}\,
\varphi(s)\,G_{in}(s,Q^2,s')\varphi(s')\;.
\label{intfin}
\end{equation}
The form factors $G_{in}(s,Q^2,s')$ are the reduced matrix
elements on the Poincar\'e group that are given by regular
Lorentz-covariant
generalized functions with test functions
$\varphi(s)\varphi(s')$.

In general, the explicit form of the functions $G_{in}(s,Q^2,s')$,
is unknown. To calculate these functions
we propose a modified impulse approximation (MIA) \cite{KrT09}.
 In contrast to the generally accepted impulse
approximation, we formulate MIA
in terms of reduced matrix elements on the Poincar\'e group (form factors)
extracted from the current matrix element and not in terms of
current operators themselves. The standard impulse approximation
is known to break the
Lorentz-covariance and the conservation law for the composite-system
electromagnetic current (see, e.g., \cite{GiG02, KeP91, KrT09}).
Note that when deriving (\ref{intfin}) we have made no assumptions about
the structure of the operator in (\ref{jc}), so that the
Lorentz-invariance and the conservation law were not broken.
MIA  means that the form factors
$G_{in}(s,Q^2,s')$
are changed for the  free two-particle
form factors $g_{0i}(s\,,Q^2\,,s')\;,\;i=C,Q,M$ of the system with no
interaction between components and with the same quantum numbers $J = S
=1\,,\,l=0$.

The free two-particle form factors also are regular
Lorentz-covariant generalized functions, so that the static limit
of, e.g.,  $g_{0i}(s\,,Q^2\,,s')$ is to be considered in the weak sense.
The result for the quadrupole form factor of our system in MIA has the
following form:
\begin{equation}
G_Q(Q^2) =\frac{2M_c^2}{Q^2}\int d\sqrt{s} d\sqrt{s'}
\varphi(s) g_{0Q}(s\,,Q^2\,,s')\varphi(s')\;.
\label{GqGRIP}
\end{equation}
The explicit form of the free two-particle quadrupole form factor
$g_{0Q}(s\,,Q^2\,,s')$ is given in Appendix.

It is easy to see that for zero values of  the parameters ($\omega_1 =
 \omega_2 =0$) of the relativistic spin rotation of the constituents in
 (A1) the free two-particle quadrupole form factor
 $g_{0Q}(s\,,Q^2\,,s')$ is zero as well as the form factor
 (\ref{GqGRIP}) of the interacting system. The existence of the
 nonzero quadrupole form factor of the system with $l=0$ is the
 consequence of the relativistic spin rotation effect.

The quadrupole moment $Q$ of the system is defined as the static limit
of the quadrupole form factor
(\ref{GqGRIP}) (see, e.g., \cite{GaO01, GiG02}):
\begin{equation}
\lim\limits_{Q^2 \to 0}G_Q(Q^2) = M_c^2\,Q\;.
\label{lim}
\end{equation}
In our case the corresponding limit is to be taken in weak sense and gives:
$$
Q = \int\limits_{2M}^\infty\frac{d\sqrt{s}}{2\sqrt{s - 4M^2}}\varphi(s)\,Q(s)\,\varphi(s) =
$$
\begin{equation}
\int\limits_0^\infty k^2\,dk\,u(k)\,Q(s(k))\,u(k)\;,
\label{Q}
\end{equation}
where   $Q(s)$ is the relativistic quadrupole-moment operator that is
the $c$-number function in the representation given by the basis
(\ref{PkJlSm}). The function $Q(s)$ has the following form:
\begin{equation}
Q(s) = -\frac{L(s)}{2M\sqrt{s}}\left(\frac{M}{\sqrt{s} + 2M} + a\right)\;,
\label{Q(s)}
\end{equation}
$$
L(s) = \frac{2\,M^2}{\sqrt{s - 4\,M^2}\,(\sqrt{s} + 2\,M)}\,\left(
\frac{1}{2\,M^2}\sqrt{s\,(s - 4\,M^2)} \right.
$$
$$
\left. + \ln\, \frac{\sqrt{s} - \sqrt{s - 4\,M^2}}{\sqrt{s} +
\sqrt{s - 4\,M^2}}\right)\;,
$$
where $a=\kappa_1 + \kappa_2\,,$ $\kappa_{1,2}$ are the anomalous
 magnetic moments of the constituents.

\section{The quadrupole moments of the $\rho$ meson and the $S$-wave
deuteron}
\label{sec: Quad rho deut}

Let  us calculate now the quadrupole moments of the $\rho $
meson and of the deuteron in the $S$-state using  (\ref{Q}), (\ref{Q(s)}) .

For the calculation of the $\rho $-meson quadrupole moment we use
the unified $\pi \& \rho$ model \cite{KrP16, KrP18} which have described
efficiently the electroweak properties of the pion and the $\rho$ meson.
In the model we have used the power-law
wave function:
\begin{equation}
u(k) = 16\sqrt{\frac{2}{7\pi b^3}}\frac{1}{(1 + k^2/b^2)^3}\;,
\label{wfPL3}
\end{equation}
with a parameter $b$.
All the
parameters of the model were fixed in \cite{KrP16}
and in the paper \cite{KrP18} the experimental value of the
$\rho$-meson magnetic moment was obtained using no additional fitting
parameters. The same values of parameters we use now to calculate the
$\rho $-meson quadrupole moment.
So, for the masses of the constituent $u$- and $\bar d$ quarks we have
$M_u = M_{\bar d} = M =0.22$ GeV and the sum of their anomalous magnetic
moments is $a = \kappa_u + \kappa_{\bar d} = 0.0268$ in quark magnetons.
The parameter of the wave function in the model (\ref{wfPL3}) is
 $b = 0.385\pm 0.019$ GeV. The result of the calculation that uses the
 formulae (\ref{Q}), (\ref{Q(s)}) with the wave function (\ref{wfPL3}) and
 parameters given above is
$Q = -0.158\pm 0.04$ GeV$^{-2}$.

Now let us consider the quadrupole moment  of the deuteron in the
$S$-state, that is without admixture of the $D$-wave in the deuteron wave
function.
We use for the
calculation the  MT wave function
\cite{KrT07MT}, that was constructed in the framework of the so-called
potentialless formulation of the inverse scattering problem
\cite{MuT81} (see also \cite{Tro94}).
The approximation for the $S$-component of the function has the form
\cite{KrT07MT}:
\begin{equation}
u(k)=\sqrt{\frac{2}{\pi}}\sum\limits_j
\frac{C_j}{(k^2+m_j^2)}\;. \label{wfMT}
\end{equation}
The parameters $C_j\;,\;m_j$ are given in the Appendix.

Other parameters entering the quadrupole moment
(\ref{Q(s)}) are well defined. The anomalous magnetic moments of the
 constituents - proton and neutron - are measured with great accuracy
\cite{PDG16}:
$a = \kappa_p + \kappa_n  =-0.1201953\pm 0.0000005$ in nuclear magnetons.
The nucleon masses are also well known \cite{PDG16}. The
 relative difference of the masses of proton and neutron is
 the value of a fraction of per cent, so we put them to be equal
 to their mean mass
 $M$ = 0.93891870 GeV. Using these parameters and the equations
(\ref{Q}), (\ref{Q(s)}) (\ref{wfMT}) we obtain the following small value
of the quadrupole moment of the $S$-wave deuteron:
$Q = -1.4\cdot 10^{-4}$ GeV$^{-2}$.

It is quite doubtful that our results for the quadrupole moments of the
$\rho $ meson and the $S$-wave deuteron could be tested experimentally in
the foreseeable future. However it seems rather interesting to us that the
quadrupole moments of so different systems are calculated in the framework
of one and the same method. Moreover, this fact encourages us in the
attempt to consider the problem of general constraints on the
quadrupole-moment values for the set of parameters considered above. The
 effectiveness of our approach for relativistic models of other
two-particle composite systems permits us to believe that such constraints
may be of rather wide validity.

\section{Properties of the relativistic quadrupole\\ moment operator}
\label{sec: Properties Q}

\begin{figure}[h!]
\epsfxsize=0.9\textwidth
\centerline{\psfig{figure=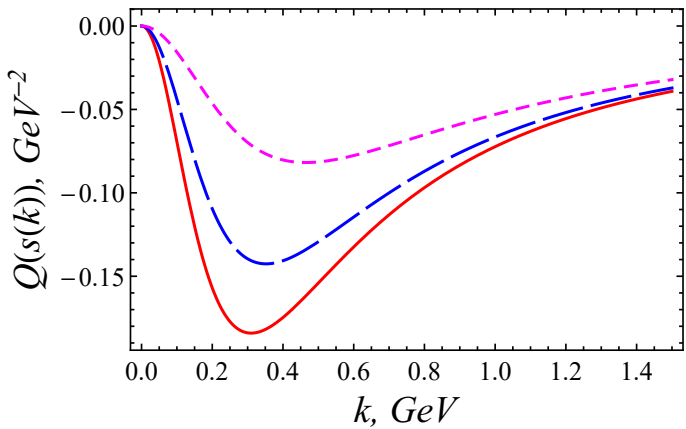,width=9cm}}
\vspace{0.3cm}
\caption{ The $c$-number quadrupole-moment operator $Q(s(k))$ (\ref{Q(s)})
 as a function of $k$ for
the sum of anomalous magnetic moments $a =0$ for
three values of the constituent mass $M$. Solid
line (red)- $M = 0.22$ GeV, dashed line (blue) - $M =
0.25$ GeV, short-dashed line (magenta) - $M = 0.33$ GeV.}
\label{fig:1}
\end{figure}

Consider the $c$-number relativistic operator of the quadrupole moment
(\ref{Q(s)}). The function in (\ref{Q(s)}) has the following
properties:
\begin{equation}
\lim\limits_{s\to 4M^2}Q(s) = 0\;,\quad \lim\limits_{s\to \infty}Q(s) = 0\;.
\label{rel}
\end{equation}
In Fig.1 the dependence of the function $Q(s(k))$ on the momentum
variable $k$ for different constituent masses and the zero value of the
sum of the anomalous magnetic moments of constituents $a=0$ is shown.
The first of the equations (\ref{rel}) means that the contribution of the
small relative momenta to the quadrupole moment is suppressed.
This means that the large-distance contribution to the relativistic
quadrupole moment is small. This contrasts fundamentally with the
nonrelativistic case (see, e.g., \cite{BrJ76}) when the nonrelativistic
quadrupole moment is defined by the wave function at large distances.
We will refer to the  relativistic wave functions that give a large value
of the probability for constituents to be found at large relative distances
as to models with weak coupling. The corresponding relativistic
quadrupole moment
(\ref{Q}) is small. The deuteron presents an example of this type of
coupling.

On the contrary, we refer to the models with the wave functions
 concentrated at small distances as to models with strong coupling of
the constituents. In such models the relativistic quadrupole-moment values
are larger.

In what follows we consider systems with weak coupling using the wave
function
(\ref{wfMT}) normalized to unity. The systems with the most strong
coupling are realized in the model with the square-law confinement.
This model with the harmonic oscillator potential is largely used
in composite quark models (see, e.g.,
\cite{BiS14}). The corresponding wave function in the
representation (\ref{PkJlSm}) with quantum numbers
$l=0\,,S=1$ is of the form:
\begin{equation}
u(k) = 2\sqrt{\frac{1}{\sqrt{\pi}b^3}}\exp{\left(-\frac{k^2}{2b^2}\right)}\;,
\label{wfHO}
\end{equation}
where the parameter $b$ determines the confinement scale.
We use also the model with intermediate coupling
(\ref{wfPL3}), that is close (see Sec. \ref{sec: Quad rho deut}) to the
model with the linear confinement \cite{Tez91}.

It follows from the conditions  (\ref{rel}) (see also Fig.1) that the
function $Q(s(k))$ has an extremum. So, one can see that the quadrupole
moment (\ref{Q}) is defined by the value of the overlap integral of the
square of the wave function and the function (\ref{Q(s)}). The largest
absolute value of the quadrupole moment is to be expected in the models
with the largest overlaps. In the strong-coupling model the position of
the maximum of the square of the wave function  (\ref{wfHO}) is defined by
the parameter $b$. There exists a value $b$ that gives the
maximum overlap and, so, the maximum value of the quadrupole moment. Our
numerical calculations confirm this statement.

The Fig.1 demonstrates also that the absolute value of the
 quadrupole-moment operator decreases appreciably with
 increasing mass of the constituents. Consequently the absolute value
 of the quadrupole moment decreases and will go to zero in the limit
 of large masses of the constituents. This is in accordance with the fact
that the quadrupole moment in the systems under consideration is a
  relativistic effect and disappears in the nonrelativistic limit.

Let us discuss now the dependence of the quadrupole moment on the sum of
anomalous magnetic moments $a$ and dislocate the important region of this
 variable for study.
For the deuteron, $a$ is known from the experiment with great
 accuracy (see Sec.\ref{sec: Quad rho deut}). For a quark-antiquark system,
 $a$ cannot be measured. However, the model independent constraints for
 the anomalous magnetic moments of $u$- and $d$- quarks were obtained
in the paper  \cite{Ger96}:
\begin{equation}
\frac{e_u + \kappa_u}{e_d + \kappa_d} = -1.77\;,
\label{rulG}
\end{equation}
where $e_{u,d}$ are the charges of  $u$- and $d$-quarks and
$\kappa_{u,d}$ are their anomalous magnetic moments (in quark magnetons).

Using the equation (\ref{rulG}) and our definition
 $\kappa_u + \kappa_{\bar d} = \kappa_u - \kappa_d = a$ it is easy to
obtain the values of anomalous magnetic moments of the quark and the
antiquark as functions of the parameter $a$. This dependence is shown in
Fig.2.

\begin{figure}[h!]
\epsfxsize=0.9\textwidth
\centerline{\psfig{figure=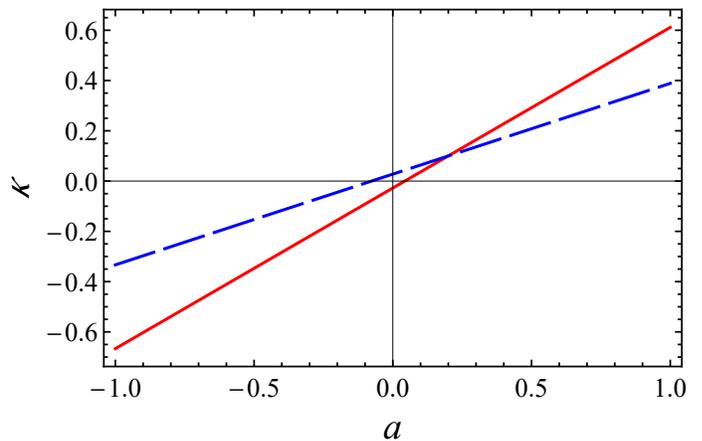,width=9cm}}
\vspace{0.3cm}
\caption{Anomalous magnetic moments of the $u$- and $\bar d$-quarks in
 quark magnetons as functions of the sum of anomalous magnetic moments $a$
corresponding to (\ref{rulG}) \cite{Ger96}. Solid line (red) -
$\kappa_u$, dashed line (blue) - $\kappa_{\bar d}$.}
\label{fig:2}
\end{figure}

For the point-like quarks ($\kappa_u = \kappa_{\bar d} = 0$), the ratio of
the magnetic moments of $u$- and $d$- quarks is exactly $-2$ that is not
far of (\ref{rulG}). The deviation of (\ref{rulG}) from this value owing to
the anomalous magnetic moments is approximately 12\% and can be considered
as a correction. So, it is natural  to consider the anomalous magnetic
moments as corrections to the point-like quark magnetic moments, too. This
allows one to consider in what follows the range of the values of the
parameter $a$ from $-0.25$ to 0.25. Fig.2 demonstrates that this interval
gives the values of anomalous magnetic moments  of the quarks that are
realistic from the point of view of the ratio (\ref{rulG}). Note that
the sum of anomalous magnetic moments of proton and neutron lays in this
interval.

The masses of the constituent $u$- and $d$-quarks are the parameters of
the composite quark model and in the current literature their values are
 always greater than 0.1 GeV. We choose the interval for the masses
 of constituents to extent from
 0.1 GeV up to 1.0 GeV. The masses of the constituents in the $\rho $ meson
and the deuteron enter this interval.

\section{Constraints on the quadrupole moment of spin one composite
system in the $S$-state of relative motion}
\label{sec: Quad constr}

Let us derive some bounds on possible values of the quadrupole moment
of two-particle systems with the total spin one in the $S$-state of the
relative motion. We consider the class of the interaction models
with the strongest coupling realized for the model with
square-law confinement
(\ref{wfHO}).

\begin{figure}[h!]
\epsfxsize=0.9\textwidth
\centerline{\psfig{figure=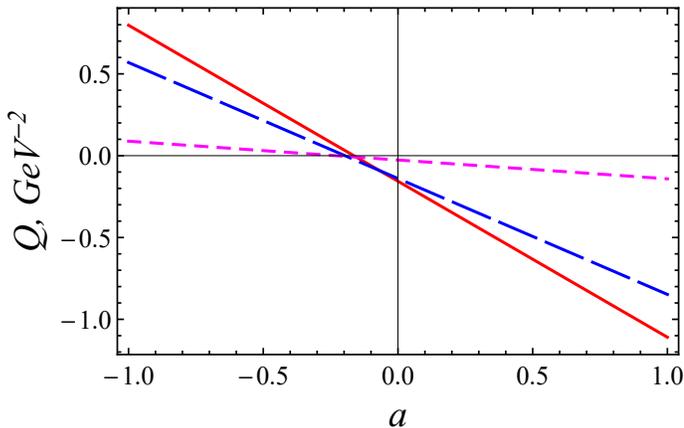,width=9cm}}
\vspace{0.3cm}
\caption{The quadrupole moment of composite system as a
function of the sum of anomalous magnetic moments of the
constituents $a$ for different model interactions at
$M=0.22$ GeV. Solid(red) line - the harmonic oscillator wave function
(\ref{wfHO}) with
$b=0.35$ GeV \cite{KrT01}, dashed line (blue) - the power-law wave
function (\ref{wfPL3}), $b=0.385$ GeV \cite{KrP18}, short-dashed line
(magenta) - weak coupling model (\ref{wfMT}) \cite{KrT07}.}
\label{fig:3}
\end{figure}

We show that in the framework of our approach, adding
physical reasoning connected with the structure of our relativistic
operator of the quadrupole moment, it occurs to be possible to obtain some
constraints for the values of the quadrupole moment of the
composite systems with quantum numbers indicated above.
 As our approach have demonstrated its effectiveness for relativistic
theory of very different two-particle composite systems \cite{KrT07, KrT01,
KrT98, KrT09prc, TrT13, KrP16, KrP18, KrN13}, it is plausible to expect
 that our constraints are of a rather general character.

The quadrupole moment of the system (\ref{Q}) is a function (see
 (\ref{Q(s)})) of three variables $Q=Q(M,b,a)$ in the case of wave functions
(\ref{wfPL3}) and (\ref{wfHO}).
In the weak coupling model (\ref{wfMT}) there is no parameter
$b$. Let us consider first the dependence of the quadrupole moment on the
sum of anomalous magnetic moments $a$  of the constituents. As can be seen
from (\ref{Q}) the quadrupole moment is a linear decreasing function of
$a$ for all models of interaction. It is plotted in Fig.3 for the models
(\ref{wfPL3}), (\ref{wfMT}), (\ref{wfHO}) with $M=0.22$ GeV. For the model
(\ref{wfHO}) we use $b = 0.35$ GeV as in pion calculations \cite{KrT01},
for the model (\ref{wfPL3}) we put $b = 0.385$ GeV used in the unified
$\pi \& \rho$ model \cite{KrP16, KrP18}. The weak-coupling wave function
(\ref{wfMT}) was normalized to unity. Fig.3 is in accordance with the
statement of Sec.\ref{sec: Properties Q} that the largest absolute value
of the quadrupole moment for the same parameters $M$ and $a$ is achieved
for the model (\ref{wfHO}).

\begin{figure}[h!]
\epsfxsize=0.9\textwidth
\centerline{\psfig{figure=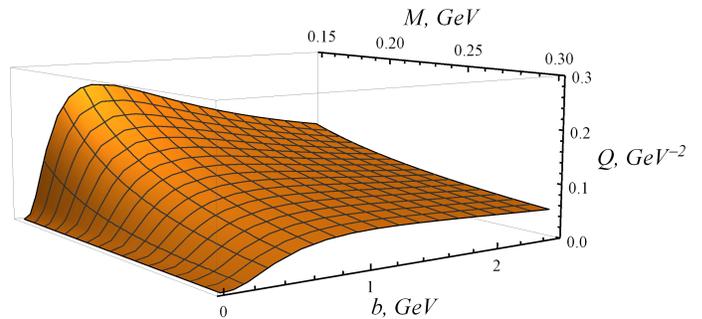,width=9cm}}
\vspace{0.3cm}
\caption{The quadrupole moment $Q(M,b,-a_m)$ (\ref{Q}) in the region
$a\le a_0,\;Q\ge 0$ (\ref{Hb}) as a function of the constituent mass $M$
and the parameter of the wave function $b$ in the model of harmonic
oscillator (\ref{wfHO}). $a_0$ is the value of the parameter $a$ for
$Q=0$. The sum of the anomalous magnetic moments  $a_m=0.25$.}
\label{fig:4}
\end{figure}

Note that a value $a=a_0$ for which the quadrupole moment is zero exists
 in all models of interaction and for all values of other parameters. This
 is due to a compensation mechanism that suppresses the relativistic
 quadrupole moment in a system with the quantum numbers $S=1,\;l=0$. This
 mechanism is caused by the existence of a structure of constituents,
namely, of the anomalous magnetic moments. The actual position of the zero
value of the quadrupole moment depends weakly on $M$ and on the choice of
the model.

First, let us consider the range of parameters that gives the
nonnegative value of the quadrupole moment:
$Q\ge$0.
Takig into account the fact that $Q$ decreases linearly with $a$ in all
cases we obtain that the upper bound in this domain
is defined by our choice of the interval for the
parameter $a$:
\begin{equation}
Q(M,b,a)\;\le Q(M,b,-a_m)\;,
\label{Hb}
\end{equation}
where $-a_m$ is the minimal admissible value. The role that plays this
parameter explains the detailed discussion in Sec.\ref{sec: Properties Q}
where we have supposed $-a_m=-0.25$.

Let us consider the dependence of the function
$Q(M,b,-a_m)$ on the parameters $M$ and $b$. In Sec. \ref{sec: Quad s-st}
analysing the structure of the operator of the quadrupole moment
(\ref{Q(s)}) we concluded that the quadrupole moment
(\ref{Q}) has a maximum at some value of the parameter $b$ in
(\ref{wfHO}), (\ref{wfPL3}). In fact Fig.4 presenting the quadrupole
moment
(\ref{Q}) as a function of parameters $M$ and $b$ in the model
(\ref{wfHO}) shows that at any fixed constituent mass
$M$ the quadrupole moment has a maximum at some $b=b_{max}(M)$. So,
the upper bound for the values of the quadrupole moment is a function of
$M$:
$Q(M,b_{max}(M),-a_m)$. The value $b_{max}(M)$ can be obtained using the
maximum condition for (\ref{Q}) for fixed value of the mass $M$ and
$a=-a_m$;
\begin{equation}
0 \le Q(M,b,a) \le Q(M,b_{max}(M),-a_m)\;.
\label{0leQ}
\end{equation}

In Sec. \ref{sec: Properties Q} we suggested, using qualitative reasonings,
that the largest value of the quadrupole moment is reached in the strong
 coupling model
 (\ref{wfHO}). The direct numerical calculation confirms this fact
 and shows that for arbitrary constituent mass the following chain of the
inequalities is valid:
$$
0 \le Q(M,b,a) \le Q(M,b_{max}(M),-a_m) \le
$$
\begin{equation}
\le Q_{HO}(M,b_{max}(M),-a_m)\;,
\label{0leQHO}
\end{equation}
where $Q_{HO}(M,b_{max}(M),-a_m)$ is the maximal value of the quadrupole
 moment in the model (\ref{wfHO}) at a fixed constituent mass..

In fact the function of mass $Q_{HO}(M,b_{max}(M),-a_m)$
gives the upper value of the quadrupole moment in our class of interaction
models. In this class the model with the square-law confinement
(\ref{wfHO}) presents the strongest coupling. The quadrupole moment
$Q_{PL}(M,b_{max}(M),-a_m)$ in the model with confinement close to the
linear one (\ref{wfPL3}) (see, e.g., \cite{KrT01}) also achieves a
maximum at some value of the model parameter $b$, the maximum value being
smaller than in the model with quadratic confinement. At the same mass the
value of maximum of $Q_{MT}(M,-a_m)$ for the weak coupling model
normalized function
(\ref{wfMT}) is even smaller. The direct calculation for
$M=0.22\,\hbox{GeV}$ gives:
$$
Q_{MT}(M,-a_m)\,<\,Q_{PL}(M,b_{max}(M),-a_m)\,<\,
$$
$$
<\,Q_{HO}(M,b_{max}(M),-a_m)\;,
$$
or actually:
$$
0.002\,\hbox{GeV}^{-2}\,<\,0.118\,\hbox{GeV}^{-2}\,<\,
$$
\begin{equation}
<\,0.120\,\hbox{GeV}^{-2}\;.
\label{QPLlQHO}
\end{equation}
The difference between the maxima for the models
 (\ref{wfPL3}) and (\ref{wfHO}) is small but the inequality (\ref{0leQHO})
is valid. The relations similar to
(\ref{QPLlQHO}) exist for all values of the constituent mass from a
chosen interval, the difference between maxima growing with mass
increasing.

\begin{figure}[h!]
\epsfxsize=0.9\textwidth
\centerline{\psfig{figure=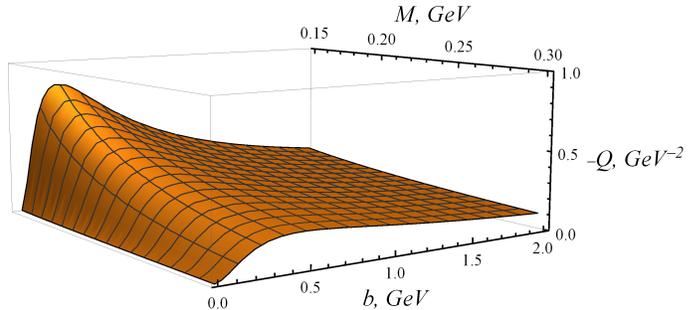,width=9cm}}
\vspace{0.3cm}
\caption{The opposite in sign value of the quadrupole moment -$Q(M,b,a_m)$
(\ref{Q}) for $a\ge a_0,\;Q\le 0$ (\ref{Lb}) as a function of
the constituent mass and of the parameter $b$ of the wave function in the
harmonic oscillator model (\ref{wfHO}).  The sum of anomalous magnetic
moments $a_m=0.25$.}
\label{fig:5}
\end{figure}

Consider now the region where
$Q\le 0$  (see Fig.3). The linear decreasing of the quadrupole moment with
increasing $a$ means that in all interaction models, the lower bound of
$Q$ is given by the largest value of $a$:
\begin{equation}
Q(M,b,a_m)\;\le\;Q(M,b,a)\;\le 0\;,
\label{Lb}
\end{equation}
where $a_m = 0.25$.

In Fig.5 the dependence of -$Q(M,b,a_m)$ on the parameters
$M$ and $b$ for the model (\ref{wfHO}) is shown.
One can  see that for an arbitrary fixed constituent mass, the function
$Q(M,b,-a_m)$ has a minimum at $b=b_{min}(M)$. The lower bound of  the
quadrupole moment is now a function of the mass $M$:
$Q(M,b_{min}(M),a_m)$.

Using reasoning and calculations analogous to those used when deriving
(\ref{0leQHO}), we estimate  the lower boundary of the quadrupole
 moment:
$$
Q_{HO}(M,b_{min}(M),a_m) \le Q(M,b_{min}(M),a_m) \le
$$
\begin{equation}
\le Q(M,b,a) \le 0\;.
\label{leQHO0}
\end{equation}
Here  $b_{min}$ is the point of
the minimal value of the quadrupole
moment in the models
(\ref{wfPL3}) and (\ref{wfHO}) at  a fixed value of
 the constituent mass, $Q_{HO}(M,b_{min}(M),a_m)$ is
the minimal value of the quadrupole
moment in the model
(\ref{wfHO}) at a fixed mass.

We can write the inequalities analogous to
(\ref{QPLlQHO}):
$$
Q_{MT}(M,a_m)\,>\,Q_{PL}(M,b_{min}(M),a_m)\,>\,
$$
$$
>\,Q_{HO}(M,b_{min}(M),a_m)\;,
$$
$$
-0.055\,\hbox{GeV}^{-2}\,> -0.384\,\hbox{GeV}^{-2}\,>\,
$$
\begin{equation}
>\,-0.395\,\hbox{GeV}^{-2}\;.
\label{QPLgQHO}
\end{equation}

The upper ($Q_{HO}(M,b_{max}(M),-a_m)$) and lower
($Q_{HO}(M,b_{min}(M),a_m)$) bounds for the values of the quadrupole
moment are shown in Fig.6 and Fig.7 as functions of the constituent mass for our
choice $a_m = 0.25$.

\begin{figure}[h!]
\epsfxsize=0.9\textwidth
\centerline{\psfig{figure=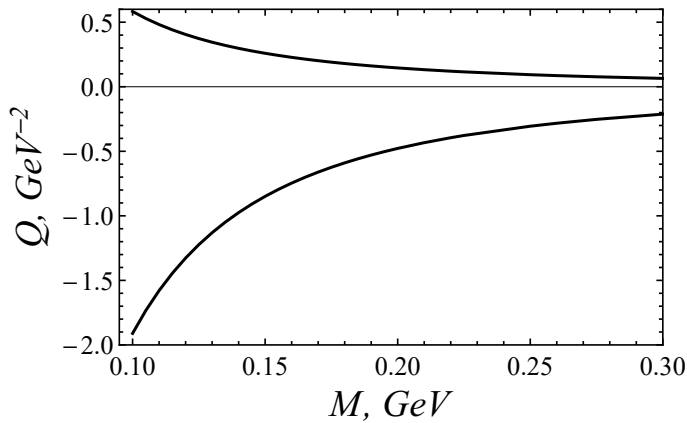,width=9cm}}
\vspace{0.3cm}
\caption{The upper and the lower bounds (\ref{0leQHO}),
 (\ref{leQHO0}) for possible values of the quadrupole
moment as functions of the constituent mass at
 $a_m = 0.25$ in quark magnetons (\ref{Hb}).}
\label{fig:6}
\end{figure}

\begin{figure}[h!]
\epsfxsize=0.9\textwidth
\centerline{\psfig{figure=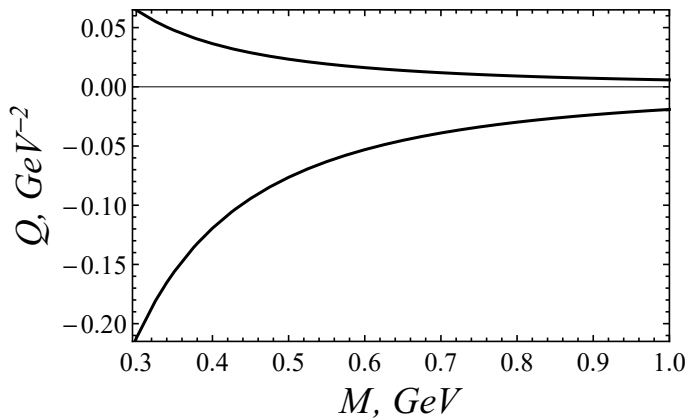,width=9cm}}
\vspace{0.3cm}
\caption{The same as in Fig.6 for other region of constituent masses.}
\label{fig:7}
\end{figure}

So, for all interaction models considered in the paper, for arbitrary
masses of constituents and for arbitrary sum of anomalous magnetic
moment from the interval $[-0.25;0.25]$, the quadrupole moment of the
two-particle system is between the curves shown in Fig.6 and Fig.7. As far as  we
know, it is for the first time that this kind of constraints is
proposed and as such it  may be ameliorated in a number of directions.
The constraints are obtained in the framework of only one approach.
However the advantages of our approach described above enable us to
believe in the validity of the constraints. The set of the interaction
models is rather limited, but we consider the interactions that are
the most popular in calculations of two-particle composite systems.
The choosen interval for the parameter $a$ plays a very important role and
the width of the band of possible values of the quadrupole moment can be
diminished efficiently for a smaller value of $a_{m}$. The detailed
discussion of the choice of its value was given above. The mass interval
considered is reasonably wide.

We compare some of results of different authors on the $\rho $-meson
quadrupole moment with our bounds. There are some values that satisfy
our constraints (see, for example, \cite{MeS02,BiS14,CaD98}) and some
others that do not.

\section{Conclusions}
\label{sec:concl}

To summarize, we construct the relativistic operator of the quadrupole
moment of two-particle composite spin one system with zero angular moment
using our version of RQM. We adopt the modified instant form  RQM that we
used previously. The derived quadrupole-moment operator in the basis with
the separated motion of the center-of-mass is a $c$-number function.

Then this operator is used to calculate, with no fitting parameters, the
values of the quadrupole moments of the $\rho $ meson ($Q_{\rho} =
-0.158\pm 0.04$ GeV$^{-2}$) and of  the $S$-wave deuteron ($Q_{d} =
-1.4\cdot 10^{-4}$ GeV$^{-2}$). The quadrupole moment of the $\rho $ meson
is obtained in the framework of the unified $\pi $\&$\rho $ model
developed by the authors in the recent papers. The quadrupole moment of
the $S$-wave deuteron is calculated using the wave function obtained by
the authors in a potentialless formulation of the inverse scattering
problem; this function has given good results for the polarization
$ed$-scattering data and for the quadrupole form factor of deuteron.

The study of the properties of the obtained quadrupole-moment
operator permits to formulate, for the first time,
the problem of the upper and lower bounds for possible values of
the quadrupole moment of a two-particle system with indicated quantum
numbers for a large range,
from $0.1$ GeV to $1$ GeV,
 of constituent masses, and to partially solve it.
The constraints are obtained in the class of interaction models for
constituents with the most strong coupling realized by the square-law
 confinement. It is shown that our limitations depend essentially on the
sum of the anomalous magnetic moments of the constituents.

\section*{Appendix}
\label{sec:Append}

The quadrupole $g_{0Q}$ form factor for free two--particle system is:
$$
g_{0Q}(s, Q^2, s') = \frac{1}{2}\,R(s, Q^2, s')\,Q^2
$$
$$
\times\left\{(s + s'+ Q^2)(G^1_E(Q^2)+G^{2}_E(Q^2))\right.
$$
$$
\times\left[\cos(\omega_1-\omega_2) - \cos(\omega_1+\omega_2)\right]
$$
$$
- \frac{1}{M}\xi(s,Q^2,s')(G^1_M(Q^2)+G^{2}_M(Q^2))
$$
$$
\times\left.\left[\sin(\omega_1-\omega_2) + \sin(\omega_1+\omega_2)\right]\right\};
\eqno{(A1)}
$$
Here
$$
R(s, Q^2, s') = \frac{(s + s'+ Q^2)}{2\sqrt{(s-4M^2) (s'-4M^2)}}\,
$$
$$
\times\frac{\vartheta(s,Q^2,s')}{{[\lambda(s,-Q^2,s')]}^{3/2}}
\frac{1}{\sqrt{1+Q^2/4M^2}}\;,
$$
$$
\xi(s,Q^2,s')=\sqrt{ss'Q^2-M^2\lambda(s,-Q^2,s')}\;,
$$
$\omega_1$ and $\omega_2$ are the Wigner rotation parameters:
$$
\omega_1 =
\arctan\frac{\xi(s,Q^2,s')}{M\left[(\sqrt{s}+\sqrt{s'})^2 +
Q^2\right] + \sqrt{ss'}(\sqrt{s} +\sqrt{s'})}\;,
$$
$$
\omega_2 = \arctan\frac{ \alpha (s,s') \xi(s,Q^2,s')} {M(s + s' +
Q^2) \alpha (s,s') + \sqrt{ss'}(4M^2 + Q^2)}\;,
$$
$\alpha (s,s') = 2M + \sqrt{s} + \sqrt{s'} $,
$\vartheta(s,Q^2,s')= \theta(s'-s_1)-\theta(s'-s_2)$, $\theta$ is
the step--function.
$$
s_{1,2}=2M^2+\frac{1}{2M^2} (2M^2+Q^2)(s-2M^2)
$$
$$
\mp \frac{1}{2M^2} \sqrt{Q^2(Q^2+4M^2)s(s-4M^2)}\;.
$$
$\lambda(a,b,c) = a^2 + b^2 c^2 -2(ab + ac + bc),$ $M$ -- the mass of a
 constituent, for example the $u$ or $\bar d$ quark or a nucleon. The
functions $s_{1,2}(s,Q^2)$ give the kinematically available region in the
plane $(s,s')$.  $G^{1,2}_{E,M}(Q^2)$-- charge and magnetic Sachs form
factors of constituents, respectively.

The ansatz for the analytic versions of the $p$-space $S$
deuteron wave function, denoted by $u(k)$, is given by (\ref{wfMT}).
In  (\ref{wfMT})
$$
m_j = \alpha + m_0\,(j-1)\;,
\eqno{(A2)}
$$
the coefficients $C_j$, the maximal value of the index
$j$ and $m_0 =0.9$ fm$^{-1}$ are defined by the condition of the best fit. One
has $\alpha = \sqrt{M\,\varepsilon_d}\;$, $\;M = 0.93891870$ GeV is average nucleon
mass, $\varepsilon_d = 2.224996\cdot 10^{-3}$ GeV is the binding energy of
the deuteron in the model \cite{KrT07}.

\begin{table}
\caption{\label{tab:coeff}Coefficients for the parametrized $S$ deuteron
wave function calculated within a dispersion approach.  The last
$C_j$ is to be computed from eq.
(A3) ($n_u$=16).}
\begin{tabular}{lrrr}
\hline
\hline
$j$&\multicolumn{1}{r}{$C_j({\rm fm}^{-1/2})$}&$j$&\multicolumn{1}{r}{$C_j({\rm fm}^{-1/2})$}\\
\hline
1\hspace{-2cm}&  0.87872995\!+\!00 &9\hspace{-2cm}&  0.59953379\!+\!07\\
2\hspace{-2cm}& -0.50381047\!+\!00 &10\hspace{-2cm}&-0.11282284\!+\!08\\
3\hspace{-2cm}&  0.28787196\!+\!02 &11\hspace{-2cm}& 0.15181681\!+\!08\\
4\hspace{-2cm}& -0.82301294\!+\!03 &12\hspace{-2cm}&-0.14519973\!+\!08\\
5\hspace{-2cm}&  0.12062383\!+\!05 &13\hspace{-2cm}& 0.96491938\!+\!07\\
6\hspace{-2cm}& -0.10574260\!+\!06 &14\hspace{-2cm}&-0.42403857\!+\!07\\
7\hspace{-2cm}&  0.59534957\!+\!06 &15\hspace{-2cm}& 0.11092702\!+\!07\\
8\hspace{-2cm}& -0.22627706\!+\!07 &16\hspace{-2cm}&\multicolumn{1}{r}{eq. (A3)}\\
\hline
\end{tabular}
\end{table}
$$
\sum\limits_{j=1}^{n_u} {C_j} = 0\;.
\eqno{(A3)}
$$

\end{document}